**Development of a Method for Tissue Elasticity Imaging Using Tagged Magnetic Resonance Imaging**

Tomoki Takeuchi, Ryosuke Nasada, and Kenya Murase*

Department of Medical Physics and Engineering, Division of Medical Technology and Science, Course of Health Science, Graduate School of Medicine, Osaka University, Osaka, Japan

*Corresponding author, Kenya Murase, Dr. Med. Sci, Dr. Eng.
Department of Medical Physics and Engineering, Division of Medical Technology and Science, Faculty of Health Science, Graduate School of Medicine, Osaka University
1-7 Yamadaoka, Suita, Osaka 565-0871, Japan
Phone & Fax: (81)-6-6879-2571,
E-mail: murase@sahs.med.osaka-u.ac.jp

Short title: Tissue elasticity imaging using tagged MRI

**Abstract**

Purpose: The intrinsic mechanical properties of soft tissues such as stiffness provide insight into a variety of disease processes. The purpose of this study was to develop a method for tissue elasticity imaging using tagged magnetic resonance imaging (MRI).

Materials and Methods: First, we developed a cyclic pressure device that used air to remotely transmit the power to generate cyclic deformation in an object in synchronization with electrocardiogram (ECG) signals. The pressure induced by the cyclic pressure device was measured by MRI-compatible force sensors consisting of optical fibers, photodiodes, and laser transmitters. Second, we developed a software to calculate Young's modulus from tagged MRI data using the harmonic phase (HARP) method and the finite element method (FEM). We also developed a software to extract tag-cross points from tagged MRI data. In the method using the HARP method, Young's modulus was calculated from the pressure data and the strain calculated by the HARP method with an assumption that the stress distribution was uniform within the object. In the method using the FEM, an FEM mesh model was firstly generated from the tag-cross points extracted from tagged MRI data and then Young's modulus was iteratively calculated using this mesh model and the displacement before and after deformation. Finally, we evaluated the usefulness and feasibility of our method using three homogeneous silicone gel phantoms with different degrees of stiffness in comparison with Young's moduli measured by a material testing machine.

Results: Our cyclic pressure device worked well without time delay between the pressure and ECG signals and air leak. The coefficient of variation of the pressure data measured by MRI-compatible force sensors was within 5%, indicating that the reproducibility of the pressure generated by our cyclic pressure device was good. The Young's moduli obtained by the material testing machine, the HARP method, and the FEM increased with increasing stiffness of the phantoms. There were relatively good agreements among the Young's moduli obtained by the three methods.

Conclusion: These results suggest that our method is useful for tissue elasticity imaging and for quantifying the stiffness of tissues.

## 1. Introduction

The intrinsic mechanical properties of soft tissues such as stiffness provide insight into a variety of disease processes. It is well known that tumors are harder than surrounding normal tissues [1–3]. Muscular dystrophy makes the muscle harder [4]. These examples indicate that the information on the stiffness has a great potential to be useful for diagnosis of such diseases. On the other hand, the virtual-reality surgical simulator has been developed [5]. The simulator needs some parameters to project human bodies in the virtual space, for example, body shape, size, weight, stiffness, and so on. As such, there exists a continuing need for the noninvasive detection, assessment, and quantification of tissue mechanical properties.

There are several methods to obtain these mechanical properties. In the field of clinical medicine, manual palpation is the most popular and simplest way of assessing tissue stiffness of patient's body. This technique, however, is limited due to its superficial and subjective nature. Recently, ultrasound elastography has become available as a new effective tool for quantifying the stiffness of soft tissues noninvasively [3]. However, a drawback of ultrasound elastography is that it is limited to imaging within a given acoustic window, cannot measure the full three-dimensional (3D) displacement of the tissue, and is highly subject to the skill and/or experience of operators.

In the field of magnetic resonance imaging (MRI), magnetic resonance elastography (MRE) has been developed to obtain information about the stiffness of tissue by assessing the propagation of mechanical waves through the tissue and has already been used clinically for the assessment of patients with chronic liver diseases [6, 7]. Although this technique is useful for quantifying the mechanical properties of soft tissues, it requires a special device for generating shear waves in the tissue and a special pulse sequence for acquiring MR images depicting the propagation of the induced shear waves. In addition, MRE requires a special software for processing the MR images of the shear waves to generate quantitative maps of tissue stiffness.

Tagged MRI is a non-invasive method for labeling magnetic tags [8–15]. The tags move following the labeled tissues. These magnetic tags, however, become blurred by longitudinal relaxation [10]. For these drawbacks, tagged MRI had not been widely used in clinical setting. However, an MRI system with higher magnetic field has advantages for performing tagged MRI [16], because the longitudinal relaxation time (T1) is prolonged. Recently, 3-Tesla (3T) MRI systems have become widely used in clinical setting. Thus, there is a possibility that the stiffness of human bodies can be easily and noninvasively measured using tagged MRI on 3T MRI systems.

There are many reports showing that the tissue strain can be obtained from deformation of tag lines obtained by tagged MRI [10, 12, 13, 17]. This suggests that tagged MRI is applicable to the measurement of elasticity in the human body.

The finite element method (FEM) is a numerical technique for finding approximate solutions to boundary value problems for partial differential equations [18]. The FEM has been applied to a variety of specializations under the mechanical engineering discipline such as aeronautical, biomechanical, and automotive industries. In a structural simulation, FEM helps tremendously in producing stiffness and strength visualizations and also in minimizing weight, materials, and costs [18].

The purpose of this study was to develop a method for tissue elasticity imaging using tagged MRI and the FEM, and to investigate its feasibility and usefulness using silicone gel phantoms with different degrees of stiffness.

## 2. Materials and methods

### 2.1. Flowchart of our tissue elasticity imaging method

Figure 1 shows the flowchart of our method for tissue elasticity imaging. The tagged MRI data were acquired in parallel with the measurement of pressure data simultaneously, and Young's moduli were calculated by the harmonic phase (HARP) method and finite element method (FEM). As shown in light gray in Fig. 1, the strain images were generated from tagged MRI data using the HARP method [10, 14, 19] with several kinds of band-pass filters. Meanwhile, the coordinates of tag-cross points were extracted to generate the two-dimensional (2D) mesh model of a phantom and a reference strain image was selected from the strain images generated using the HARP method. Then, Young's modulus was calculated using the strain images and pressure data.

In the method using the FEM shown in dark gray in Fig. 1, tag-cross points were firstly extracted from the tagged MRI data at the first frame, from which the FEM mesh model was generated. Next, the displacement was calculated by assuming a temporary Young's modulus and was compared with that obtained from the tagged MRI data at frame *t*. Finally, Young's modulus was obtained by repeating the above calculations with the modified values of Young's modulus until the difference between the displacements obtained by the FEM and tagged MRI became smaller than that set beforehand.

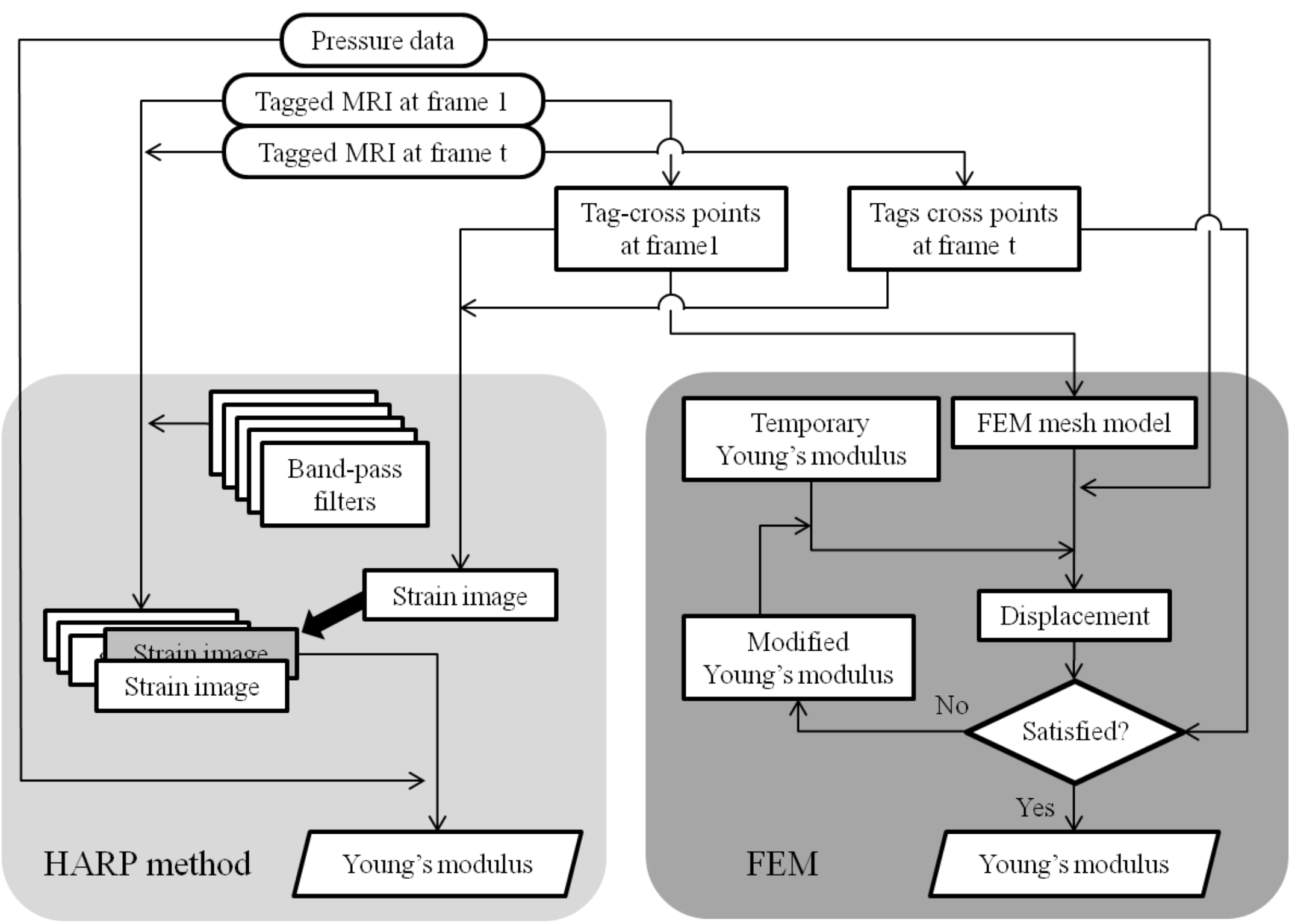

**Figure 1 Flowchart of our method for tissue elasticity imaging.**

### 2.2. Experimental apparatus

#### 2.2.1. Phantoms

For phantom studies, we made three homogeneous cylindrical silicone gel phantoms (70 mm in diameter and 40 mm in height) with different degrees of stiffness. They were named alphabetically as A, B, and C in an increasing order of stiffness, *i.e.*, Phantom A is the softest and Phantom C is the hardest. Figure 2 shows a photograph of Phantom A.

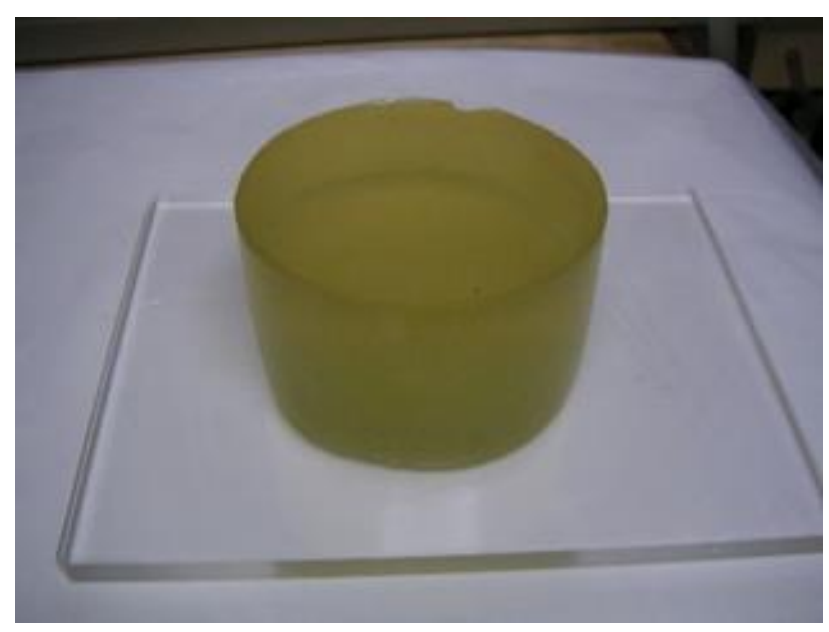

**Figure 2 Photograph of a homogeneous silicone gel phantom (Phantom A).**

Young's moduli of the phantoms were measured using a material testing machine (EZ-TEST 100N, Shimadzu Co., Japan). First, we measured the relationship between the applied pressure and the displacement of the phantom, and then calculated Young's modulus from these applied pressure and measured displacement values.

### 2.2.2. Cyclic pressure device and MRI-compatible force sensor

The cyclic deformation of an object and synchronization with MRI are indispensable for our method. Although many kinds of actuators for generating cyclic deformation are on sale, most of them are made from metals that are not suitable for MRI. Then, we made a cyclic pressure device without any paramagnetic materials, which can be operated in an MRI gantry. The cyclic pressure device we made is shown in Fig. 3. It employed air to remotely transmit the power to press an object in the MRI gantry. This device consisted of two main parts. One is an air pump [Fig. 3(a)] and the other is a cyclic pressure device [Fig. 3(b)]. The air pump has an interface to communicate with a personal computer (PC), by which we can control the frequency of cyclic pressure and the quantity of air to be pushed out. In addition, the air pump has an electrocardiogram (ECG) generator that can generate electric pulse signals to synchronize with MRI data acquisition. The ECG signals were output to PC through AD converters, and were stored in the PC with a software (C-LOGGER, CONTEC Co., Ltd., Japan).

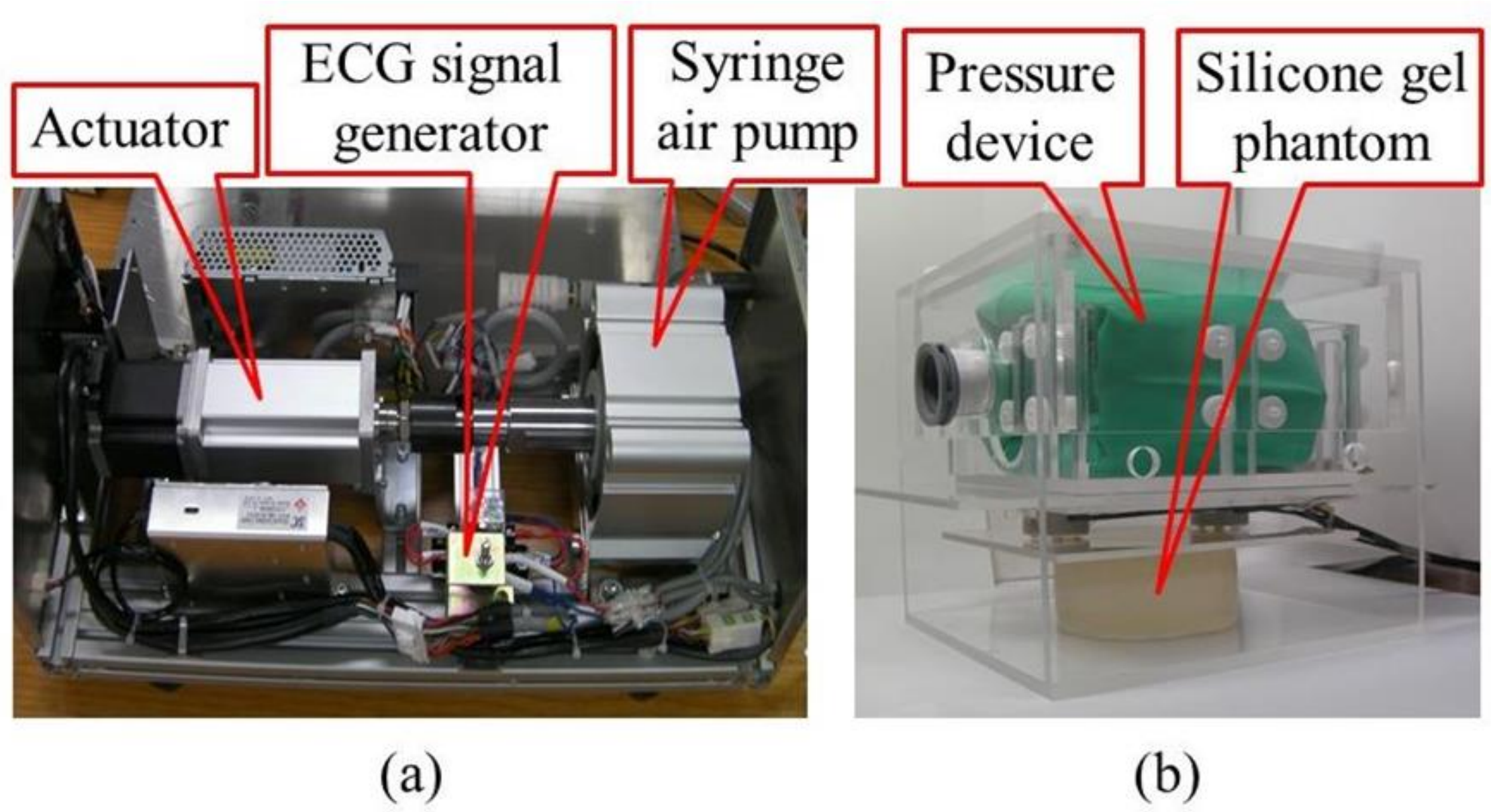

**Figure 3 (a) Photograph of an air pump to push air out of the pump under the control by a personal computer (PC). (b) Photograph of a cyclic pressure device and a silicone gel phantom.**

Figure 4 illustrates our system for tissue elasticity imaging. As shown in Fig. 4, the tagged MRI and pressure data were acquired simultaneously. The air pump and PC were placed outside the 5-gauss line to minimize the interference from the magnetic field in MRI.

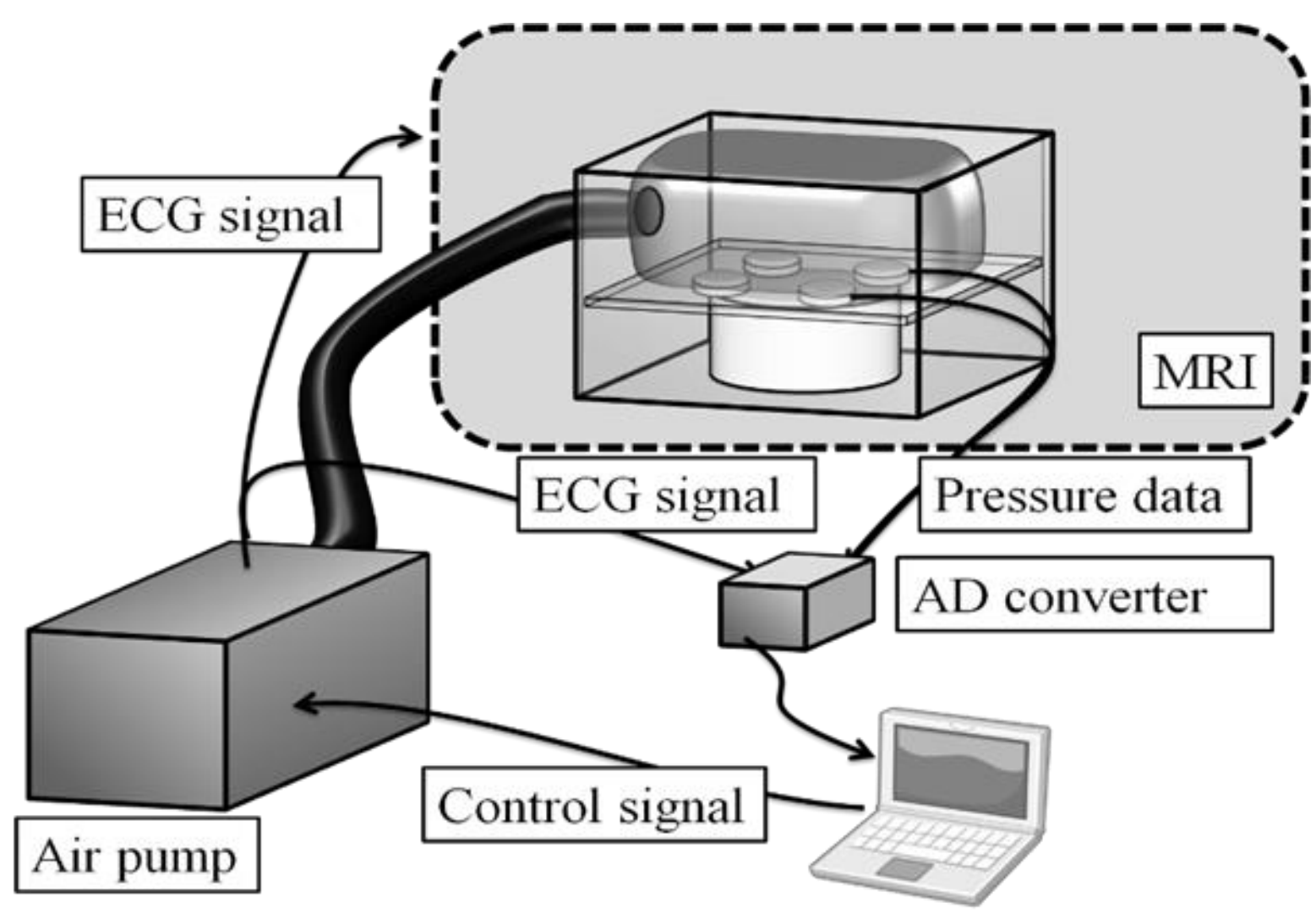

**Figure 4 Illustration of our system for tissue elasticity imaging. The tagged magnetic resonance imaging (MRI) and pressure data are acquired simultaneously through electrocardiogram (ECG) signals under the control by a PC.**

The air pump has a hose to actuate the cyclic pressure device and three electric lines for communicating with PC and for outputting pulse signals to PC and MRI. The cyclic pressure device, MRI-compatible force sensors, and a phantom were setup as illustrated in Fig. 4. They were placed in the MRI gantry.

Since the length of the hose between the air pump and the cyclic pressure device is approximately 6 m, this may cause a time lag between the pressure cycle and ECG signal. In addition, if there is an air leak from the junction of the hose, it is difficult to transmit the power correctly to press an object. Then, we checked whether there were a time lag between the pressure cycle and ECG signal and air leak by measuring the pressure and ECG signals for 500 cycles.

The pressure produced by the air pump was measured using the MRI-compatible force sensor shown in Fig. 5(a) [20, 21], which was placed between the cyclic pressure device and an object (silicone gel phantom) [Fig. 3(b)]. In this study, two force sensors were used. These force sensors consisted of power amplifiers having photodiodes and laser transmitters and sensor heads sensing applied forces. The measured data were output to PC through AD converters, and were stored in PC with a software (C-LOGGER, CONTEC Co., Ltd., Japan) as in the case of the ECG signals. It should be noted that the force sensors were calibrated by measuring the relationship between known applied forces and resultant output voltages of the photodiode beforehand [20, 21].

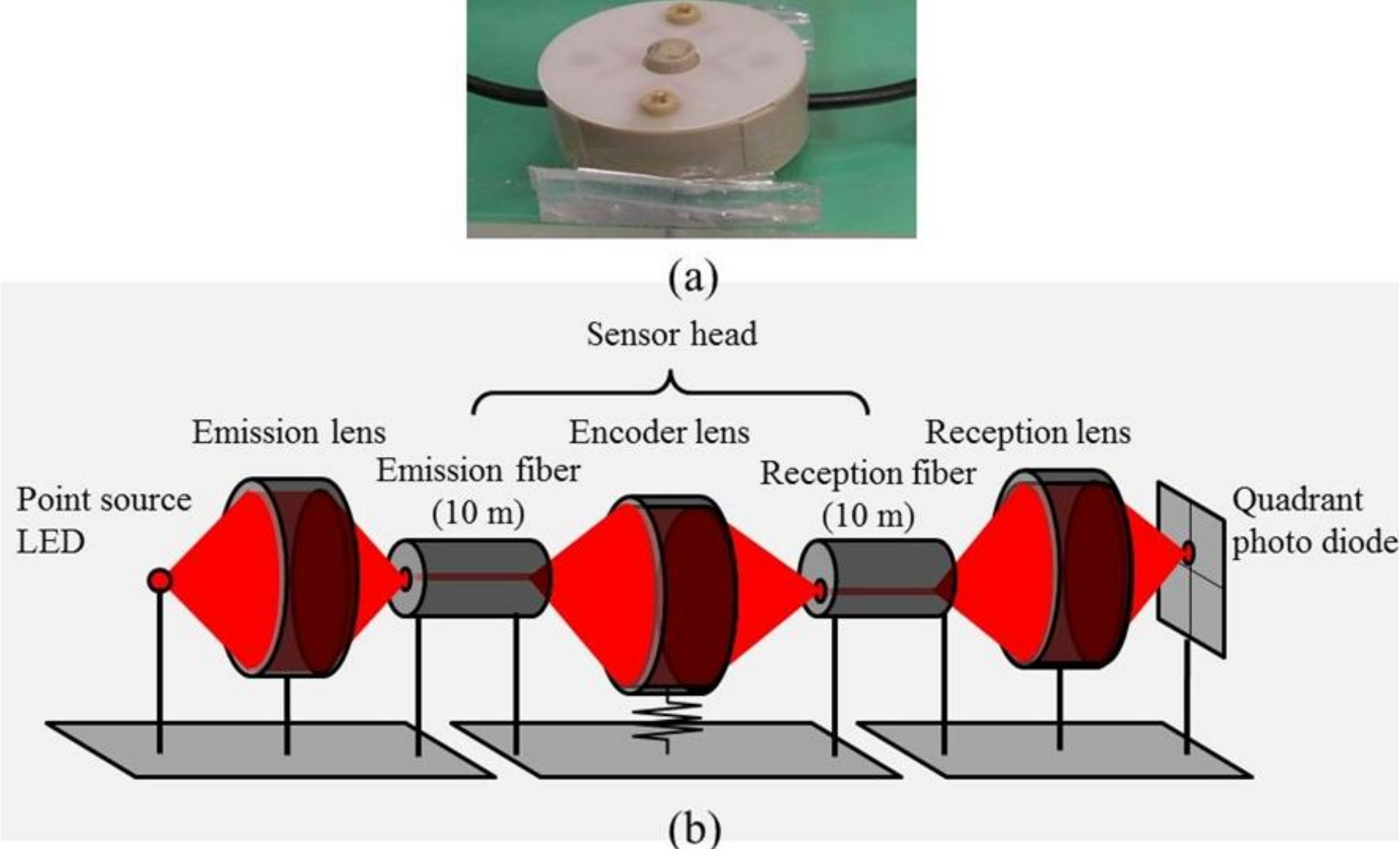

**Figure 5 (a) Photograph of an MRI-compatible force sensor head. (b) Illustration of the mechanics of an MRI-compatible force sensor head.**

The schematic diagram of an actual implementation of our MRI-compatible force sensor head [20, 21] is shown in Fig. 5(b). We employed two optical lenses to connect the optical devices and the optical fibers transmitting the focused image position. Light signal emitted from the point-source red light-emitting diode (LED) (VS679TM,

Alpha-One Electronics Co., Japan) throws its image on the end face of the multi-core emission fiber (FU-77, Keyence Co., Japan) through the emission lens (W18-S0290-063-ABC, Nihon Sheet Glass Co., Japan). The light emitted from the other end of the emission fiber is condensed again by the encoder lens (the same model as the emission lens) and throws light distribution on the end face of the reception fiber (the same model as the emission fiber). The light distribution is transmitted through the reception fiber, and is finally focused onto the imaging plane of the quadrant photo diode (MI-1515H-4D, Moririka Co., Japan) through the reception lens (the same model as the emission lens). Both the emission and the reception optical fibers are 10 m long and have 217 cores in diameter of 1.3 mm.

### 2.3. Imaging procedure

Tagged MRI data were acquired using a 3T MRI system (SIGNA EXCITE HD 3.0T, GE Medical Systems, Inc., USA). The scan parameters for tagged MRI were as follows: tag pattern = grid; tag space = 3 pixels; tag angle = 0 degree; field of view = $160\times160$ $mm^2$; matrix size = $256\times256$ pixels; number of frames in a pressure cycle = 20; number of excitations = 1; flip angle = 12 degrees; and slice thickness = 8 mm. Note that the tag angle is defined as 0 degree when the tag line is horizontal.

Eight-channel phased array coil was used and the pulse sequence for cardiac studies in the clinical setting was used for acquiring tagged MRI data. This pulse sequence is based on the complementary spatial modulation of magnetization (CSPAMM) tagging pulse sequence [16, 22]. CSPAMM acquires image data after eliminating the longitudinal magnetization at arbitrary locations. Thus, the tagged regions in the image have no signals. 2D-SPAMM is the pulse sequence made by extending the above CSPAMM to two dimensions, by which the grid-like tag lines can be obtained [10].

### 2.4. Automatic extraction of tag-cross points

Tag-cross points were used as the feature points for tracking deformation. To find these tag-cross points is time-consuming and labor-intensive. Then, we developed a software for automatic extraction of the tag-cross points. Figure 6(b) shows an example of the 3D-surface display of the tagged MR image shown in Fig. 6(a). In our software, it was assumed that the tag lines were located at the bottom of the curve approximated as a parabola shown in Fig. 6(c), and it was necessary to input the position of starting point and tag angle. Our software detected the same bottom line using the different starting points located near the bottom line.

We explain the algorithm to extract tag-cross points in more details in the following. As an example, we consider the column array of N×M shown in Fig. 6(d).

(1) First, we give the coordinate of a starting point and tag angle as $(n_0, m_0)$ and 0, respectively.

(2) Second, we search for the coordinate of the pixel whose brightness is the minimum of those at three pixels $(n_0 + 1, m_0 + 1)$, $(n_0, m_0 + 1)$, and $(n_0 + 1, m_0 + 1)$, and name this coordinate as $(n_{min}, m_0 + 1)$.

(3) Third, we represent the brightness at $(n_{min} - 1, m_0 + 1)$, $(n_{min}, m_0 + 1)$, and $(n_{min} + 1, m_0 + 1)$ as $y_{n-1}$, $y_n$, and $y_{n+1}$, respectively, and calculate the coordinate of the bottom as follows:

$$\begin{cases} y_{n-1} = a(x_{n-1} - b)^2 + c \\ y_n = a(x_n - b)^2 + c \\ y_{n+1} = a(x_{n+1} - b)^2 + c \end{cases} \tag{1}$$

If we assume that $x_{n-1} = -1$, $x_n = 0$, and $x_{n+1} = 1$, Eq. (1) is reduced to

$$\begin{cases} y_{n-1} = a(1 + b)^2 + c \\ y_n = ab^2 + c \\ y_{n+1} = a(1 - b)^2 + c \end{cases} \tag{2}$$

Solving Eq. (2). we obtain the coordinate of the tag line ($b$) as

$$b = \frac{y_{n-1} - y_{n+1}}{2(y_{n-1} + y_{n+1} - 2y_n)} \tag{3}$$

(4) Fourth, we repeat the above procedures (2) to (4) by taking $(n_{min}, m_0 + 1)$ as the next starting point given by $(n_1, m_1)$ until the new starting point arrives at the boundary of a region of interest (ROI).

(5) Fifth, we repeat the above procedures to the opposite direction from the first starting point. The tag angle of the opposite direction is regarded as $\pi$. In this case, the first starting point $(n_0, m_0)$ may not be located on the tag line. Therefore, one of the extracted coordinates from (1) to (4) is set as a new starting point.

(6) If we complete the above procedures (1) to (5), we can extract one tag line. We then specify a new starting point and repeat the above procedures until all tag lines for a tag angle of 0 degree are extracted.

(7) If we repeat the above procedures (1) to (6) for tag angles of $\pm\pi/2$ and find the points at the intersection among tag lines, we can obtain tag-cross points.

To investigate the accuracy of the above algorithm, we compared the tag-cross points extracted by the above algorithm and those obtained manually.

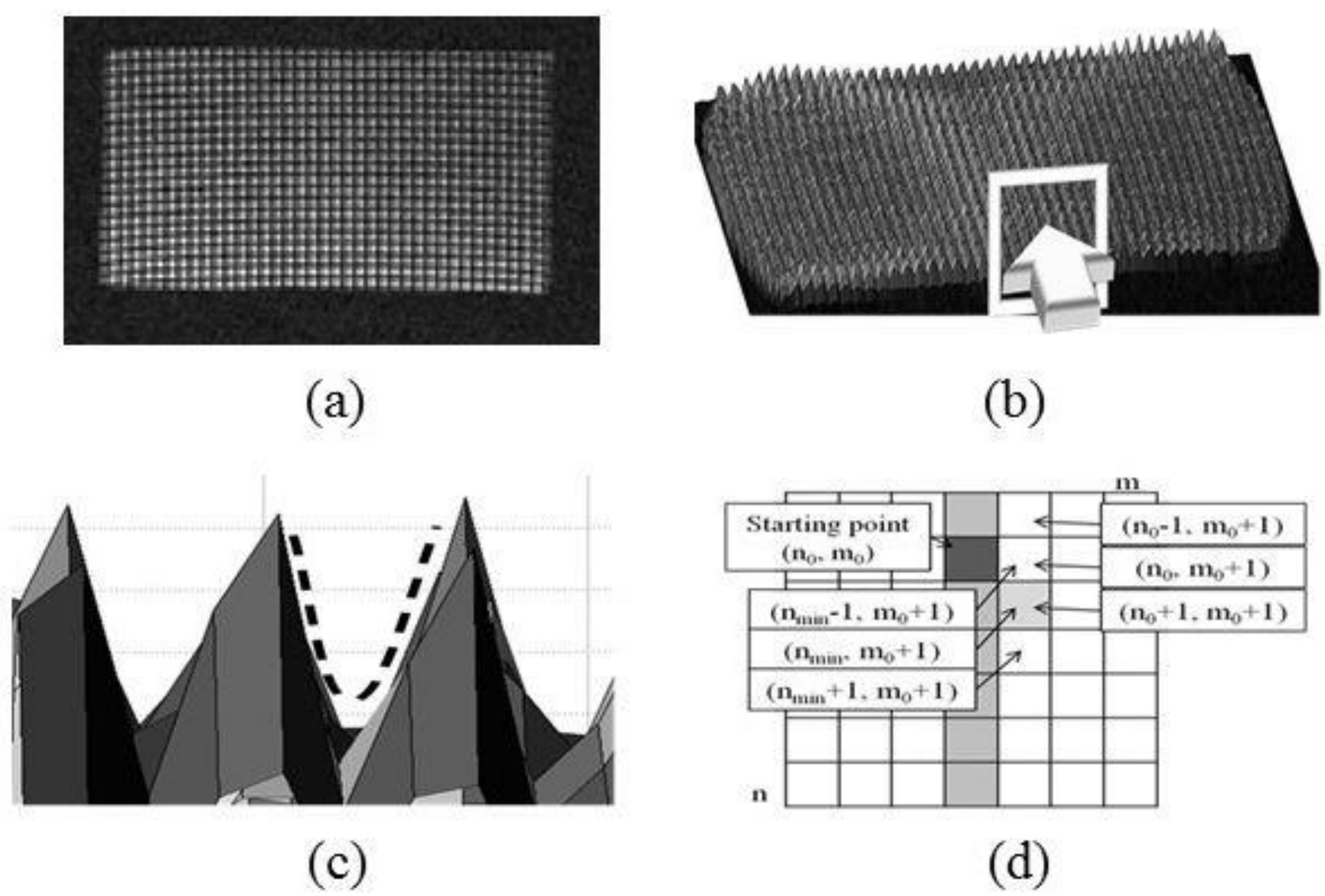

**Figure 6 (a) Example of the tagged MR image of a homogeneous silicone gel phantom. (b) Three-dimensional (3D)-surface display of the tagged MR image shown in (a). (c) Enlarged view of the area shown by white rectangle in (b), which is observed from a direction shown by arrow in (b). Note that one of the valley-shaped regions formed by tag lines was approximated by parabola shown by dashed line in (c). (d) Notations used for explaining a method for extracting tag lines and tag-cross points.**

### 2.5. Harmonic phase (HARP) method

The HARP method is a method for extracting the form of tag lines from tagged MR images [8, 9]. The details of the HARP method are described in Appendix. Figures 7(a) and 7(b) show an example of the tagged MR image of a homogeneous silicone gel phantom that has horizontal and vertical tag lines and its Fourier-transformed image, respectively. As shown in Fig. 7(b), nine peaks are seen in the frequency domain. The peak at the center in the frequency domain is a direct current (DC) peak. The upper and lower peaks are the harmonic peaks derived from the horizontal tag lines, whereas the right and left peaks are derived from the vertical tag lines. It should be noted that image analyses for the horizontal and vertical tag lines were performed separately in this study. The detailed procedure of image analysis for the horizontal tag lines is described in Appendix.

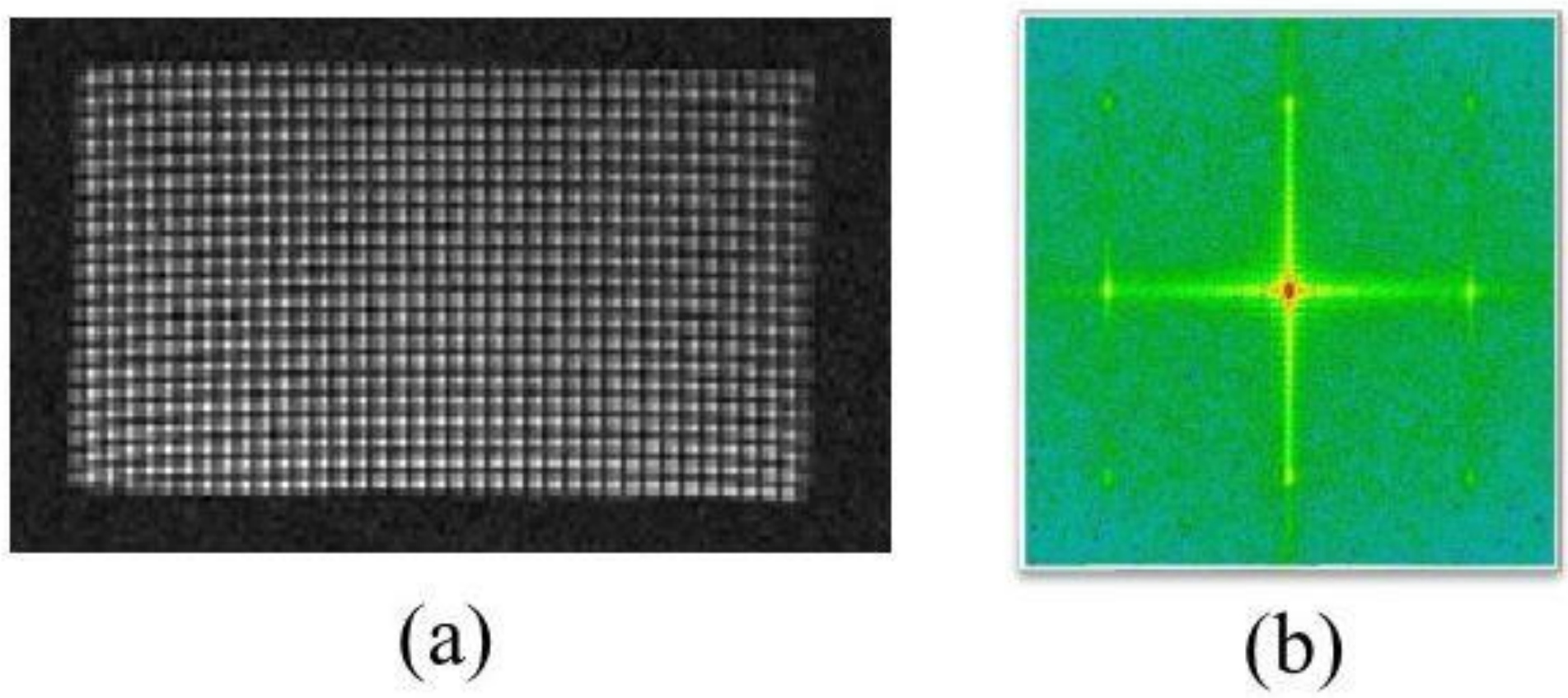

**Figure 7 (a) Example of the tagged MR image of a homogeneous silicone gel phantom. (b) Magnitude image of the Fourier-transformed image of (a) in the frequency domain.**

Young's modulus ($E$) is defined as a ratio of the stress (σ) to the strain (ε), *i.e.*, $E = \sigma/\varepsilon$. Thus, it is possible to calculate Young's modulus from the pressure data measured by MRI-compatible force sensors (Fig. 5) and the strain obtained by the HARP method (see Appendix).

### 2.6. Finite element method (FEM)

As previously described, the FEM is a numerical technique that gives approximate solutions to differential equations that model problems arising in physics and engineering [18]. As in simple finite difference schemes, the FEM requires a problem defined in geometrical space to be subdivided into a finite number of smaller regions called "mesh". The detailed explanation of the FEM is given in [18].

In our FEM software, the FEM mesh model, external force data, Poisson's ratio, and Young's modulus were used as input data. The FEM mesh model was generated from the tag-cross points obtained by tagged MRI (Fig. 6), and the external force data were measured by MRI-compatible force sensors (Fig. 5). In this study, we assumed that the phantoms were incompressible, and Poisson's ratio was taken as 0.49 in our

FEM software.

Our FEM software can calculate the strain and stress distributions within an object when Young's modulus and external forces are known. On the other hand, when Young's modulus is unknown and the displacements and external forces are known, Young's modulus can be calculated by iterative calculations with assumption of various values of Young's modulus.

Our FEM software calculates Young's modulus as follows. First, our FEM software calculates the displacements and the coordinates of nodes using an appropriate Young's modulus. Because the actual displacement is known from the tag-cross points obtained by tagged MRI as previously described, the distances between the actual nodes and the nodes calculated by our FEM software can be obtained. Next, we repeat these calculations using modified Young's modulus and continue these calculations until the distances between the actual and calculated nodes become minimum.

Since the FEM is generally based on the premise that the internal shape of the object after deformation is known, it is not applicable to the case when the internal shape of the object after deformation is not known. However, the internal shape of the object after deformation can be obtained by using tagged MRI. Thus, Young's modulus within the object can be obtained by combining FEM with tagged MRI. In this study, Young’s modulus in each homogeneous silicone gel phantom was obtained using the method mentioned above using the FEM mesh model generated from tag-cross points.

To investigate whether our FEM software is applicable to the case when Young's modulus is not uniform, we performed simulation studies as follows. First, we generated the FEM mesh model from tag-cross points obtained by tagged MRI. The Young’s modulus actually measured by a material testing machine (EZ-TEST 100N, Shimadzu Co., Japan) was used as that of this model and Poisson's ratio was assumed to be 0.49. The external forces actually measured by MRI-compatible force sensors (Fig. 5) were used as input data to our FEM software.

## 3. Results

### 3.1. Cyclic pressure device and MRI-compatible force sensor

Figure 8 shows the force values measured by MRI-compatible force sensors 1 and 2 as a function of time together with ECG signals. As shown in Fig. 8, there was no time delay between the pressure and ECG cycles, indicating that although the air pump is approximately 6 m away from the cyclic pressure device as previously described, the cyclic pressure device and ECG signal generator were synchronized correctly.

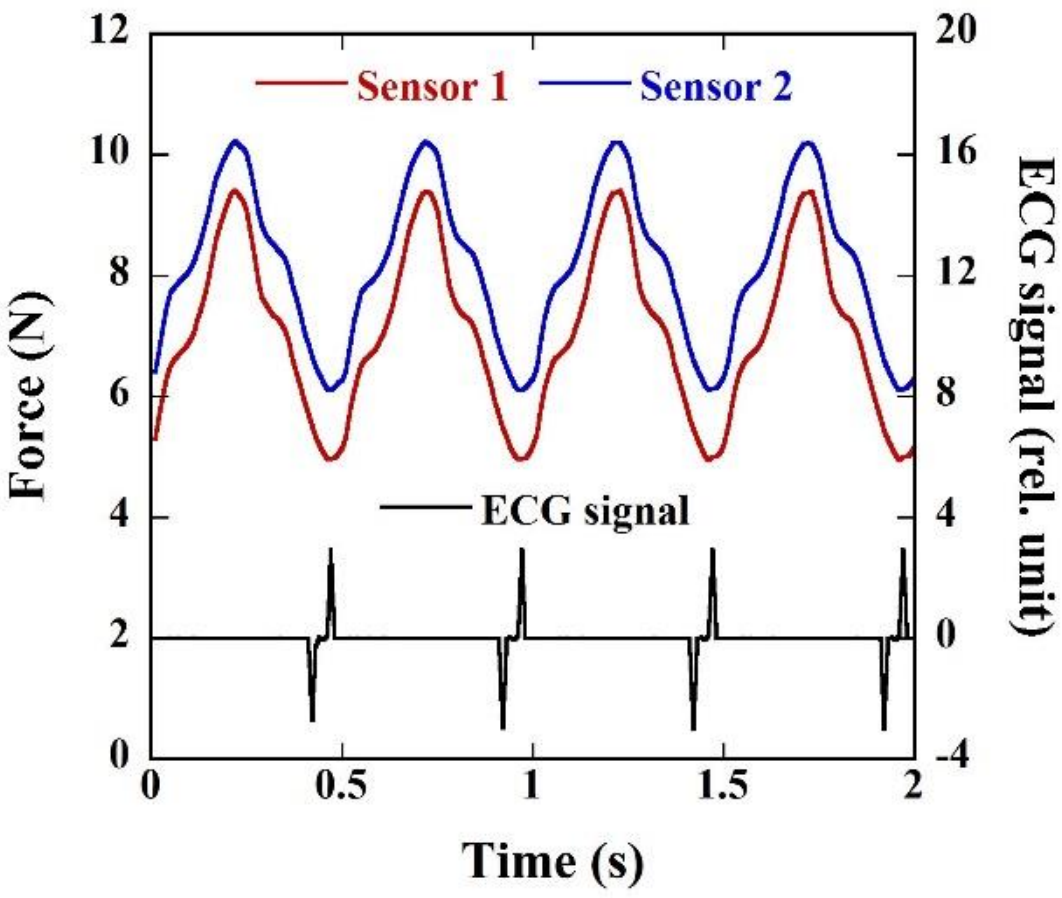

**Figure 8 Pressure data measured by MRI-compatible force sensors 1 and 2 and ECG signals as a function of time.**

Figure 9 shows the mean ± standard deviation (SD) of the force values measured by MRI-compatible force sensors 1 and 2 for 500 cycles. The coefficient of variation (CV) was 4.79 ± 0.49% and 3.89 ± 0.40% for sensors 1 and 2, respectively. As shown in Fig. 9, the SD and CV values were small, indicating that the reproducibility of these force sensors and the pressure cycles generated by the cyclic pressure device was good and there was no air leak from the junction of the hose between the air pump and the cyclic pressure device.

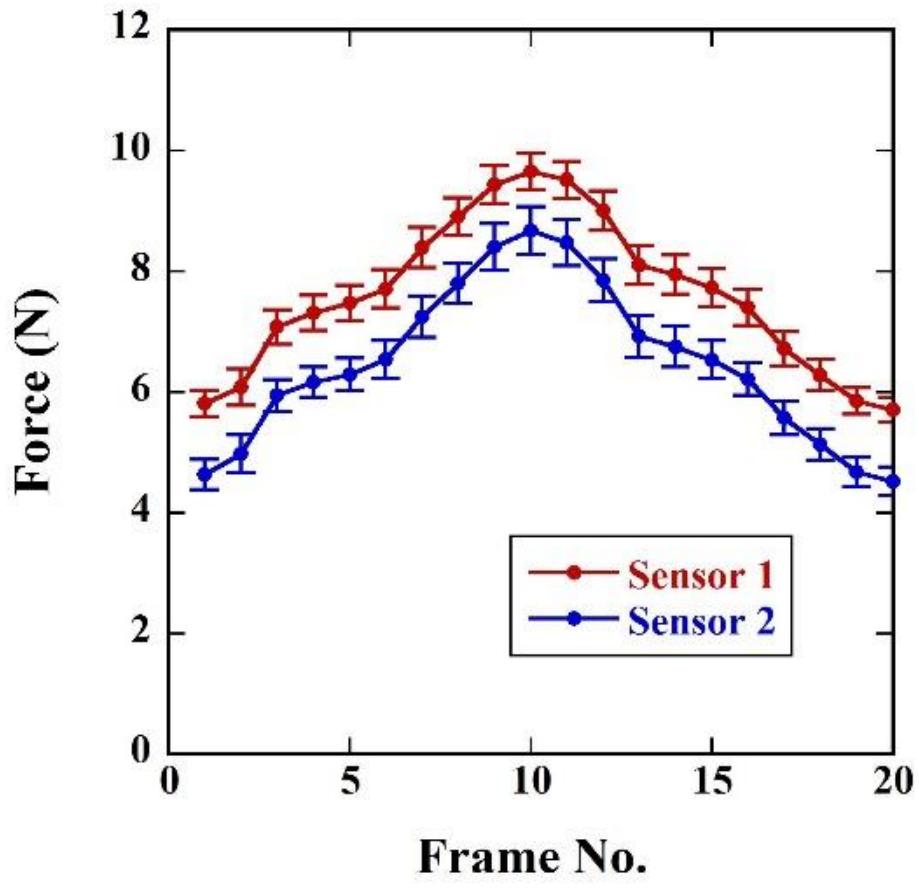

**Figure 9 Force values measured by MRI-compatible force sensors 1 and 2 as a function of frame. Data are represented as mean ± SD for 500 cycles.**

### 3.2. Automatic extraction of tag-cross points

Figure 10(a) shows the tag-cross points obtained by our software for automatic extraction of tag-cross points. Although the tag-cross points inside the phantom could be extracted accurately, some false points were observed near the boundary of the phantom. The tag-cross points after correcting these false points manually are shown in Fig. 10(b). The coordinates of these tag-cross points were used as nodes in the FEM mesh model.

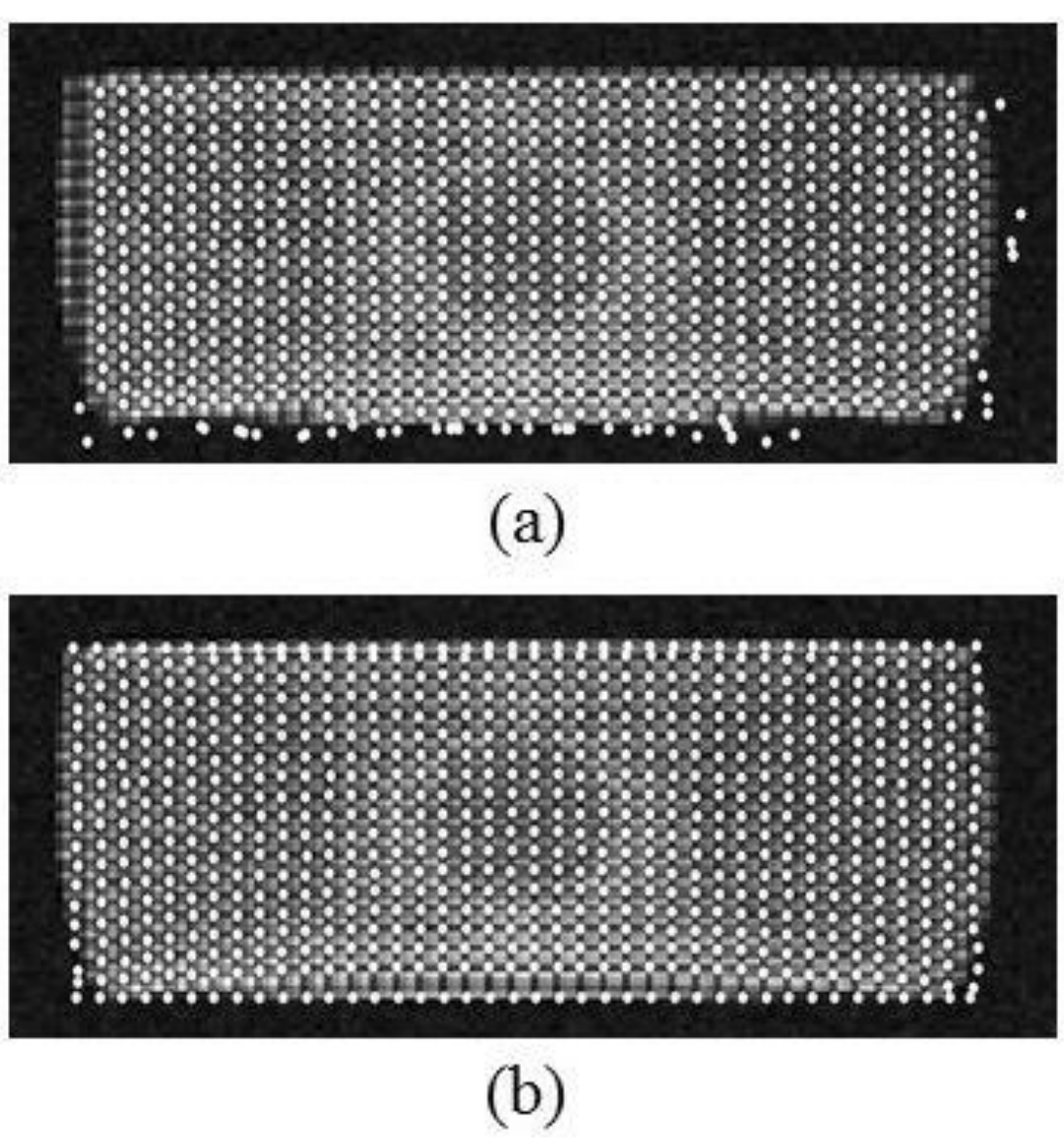

**Figure 10 (a) Example of tag-cross points extracted automatically using our software. (b) Tag-cross points after manual correction.**

### 3.3. HARP method

#### 3.3.1. Correction of phase wrap

Figure 11 shows the results of phase-wrap correction using the linear-interpolation method for a homogeneous phantom. Figures 11(a) and 11(b) show the tagged MRI and phase angle images, respectively. Figure 11(c) shows the profile of the phase angle $[I_{\varphi}(\mathrm{t})$ given by Eq. (A2)] along the white line shown in Fig. 11(b). Figure 11(d) shows the profiles of the gradient of phase angle $[\partial I_{\varphi}(t)/\partial \mathrm{y}$ given by Eq. (A5)] without (closed circles) and with correction of phase wrap (open squares). Figures 11(e) and 11(f) shows the gradient image of phase angle without and with wrap correction, respectively. As shown in Fig. 11(d) (closed circles), when the phase wrap was not corrected, the gradients of phase angle had negative values. On the other hand, when the phase wrap was corrected using the linear-interpolation method, the gradients of phase angle became almost constant [open squares in Fig. 11(d)] and the image of the gradient became almost uniform [Fig. 11(f)], suggesting that our method is effective for correction of phase wrap.

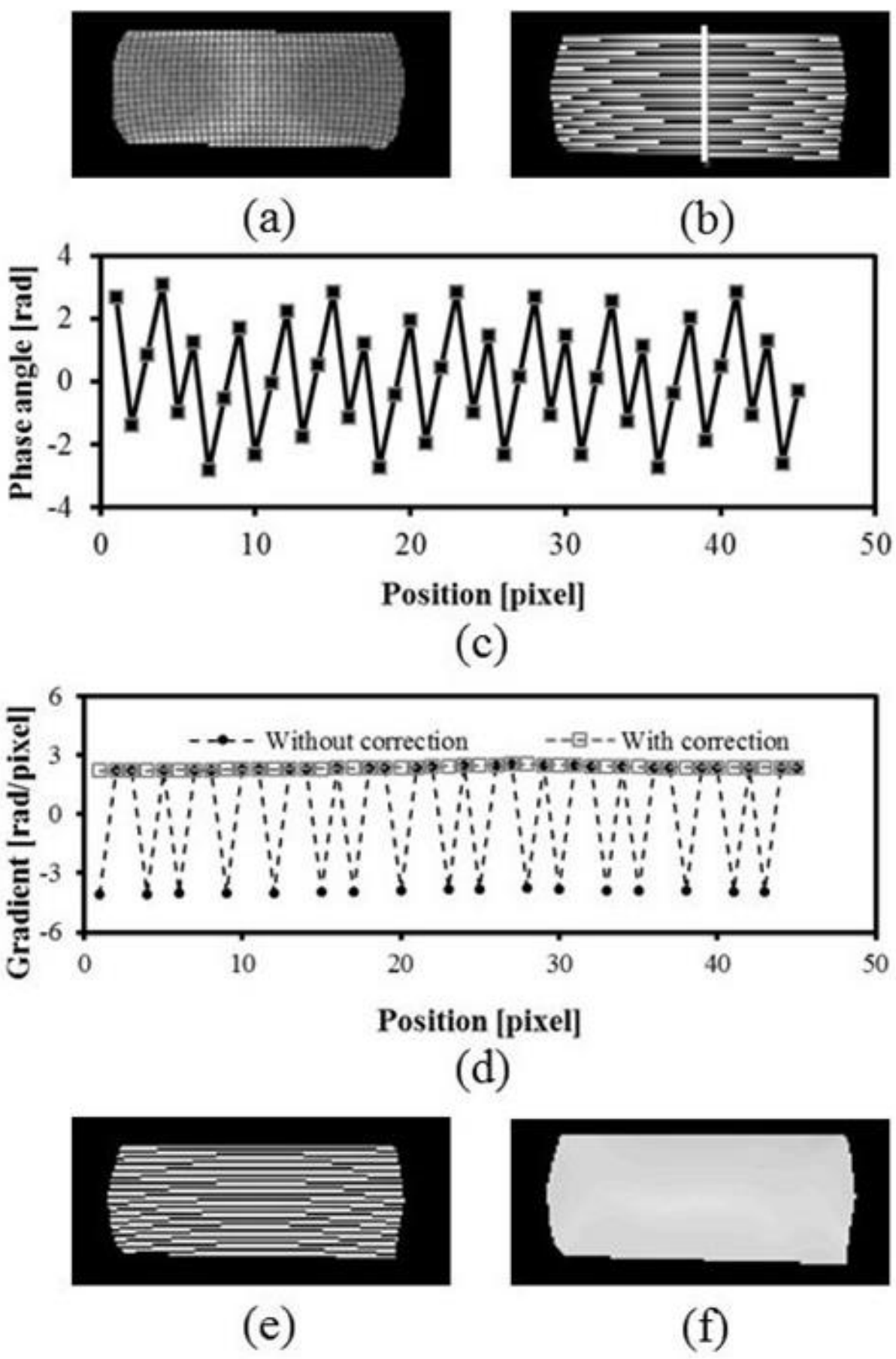

**Figure 11 (a) Tagged MRI of a homogeneous phantom. (b) Phase angle image $[I_\varphi(t)]$. (c) Profile along the white line shown in (b). (d) Gradient of phase angle. Uncorrected gradient line has several negative values due to phase wrap. These negative values were removed by the linear-interpolation method. (e) Gradient image of phase angle $[\partial I_\varphi(t)/\partial y]$. (f) Wrap-corrected gradient image of phase angle.**

### 3.3.2. Strain image generated using the HARP method

Figures 12(a), 12(b), and 12(c) show the gradients of phase angle along the white line shown in Fig. 12(c), calculated after phase wrap correction with (circles) and without pressure (triangles) as a function of position, the strain obtained from the calculated gradients of phase angle with and without pressure using Eq. (A6), and the strain image generated by using Eq. (A6) pixel by pixel for a homogeneous phantom, respectively. As shown in Fig. 12, the strain could be obtained from the gradients of phase angle using Eq. (A6).

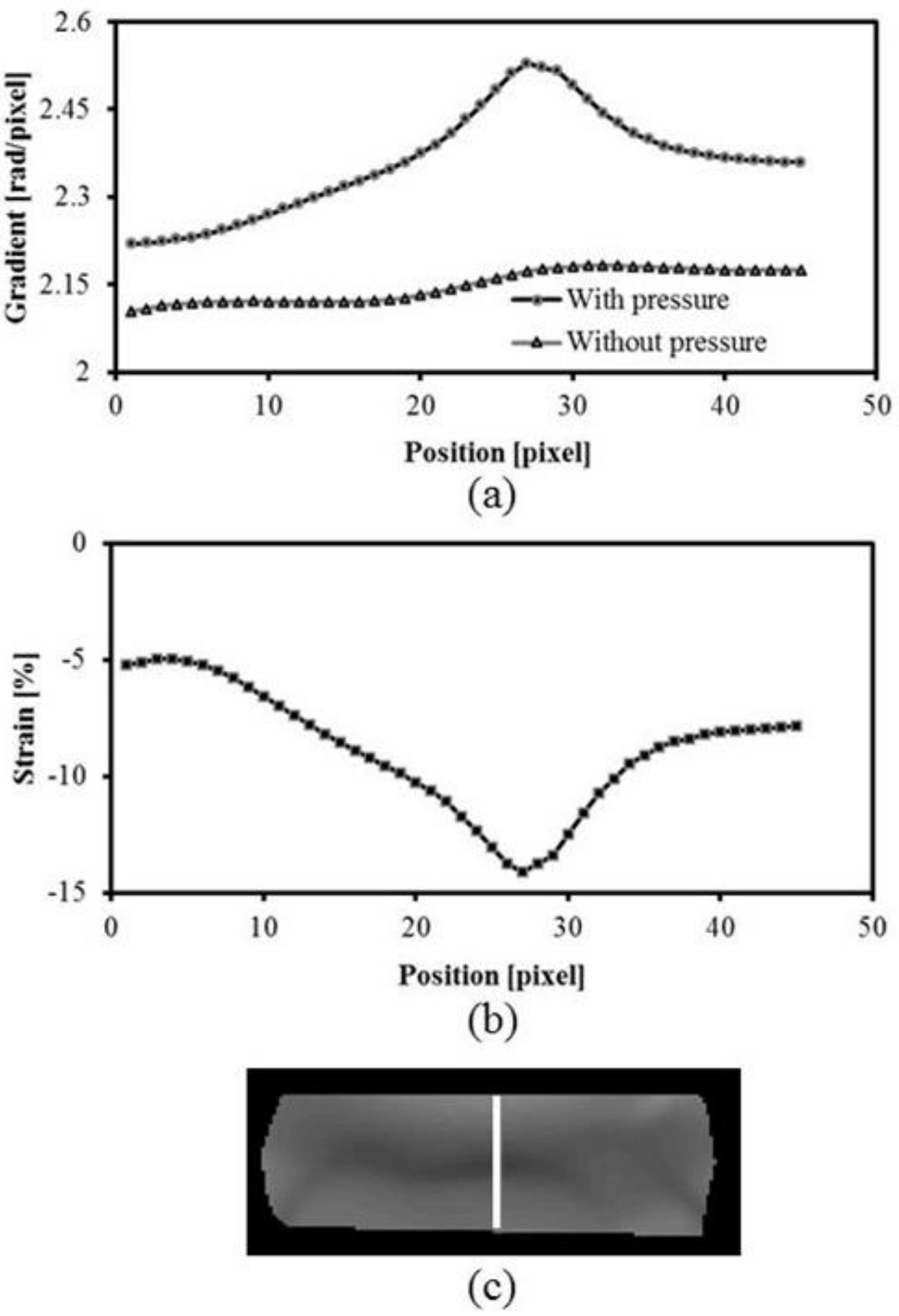

**Figure 12 (a) Calculated gradients of phase angle with and without pressure as a function of position in a homogeneous phantom. (b) Strain obtained from the calculated gradients of phase angle with and without pressure using Eq. (A6). (c) Strain image generated by using Eq. (A6) pixel by pixel.**

### 3.4. FEM

#### 3.4.1. Calculation of Young's modulus using the FEM

As shown in Fig. 13, the relationship between the index and Young's modulus had a minimum value in all phantoms, from which Young's modulus could be obtained. The index value is given by $\sqrt{\sum_{i=1}^{N} d_i^2/N}$, where $d_i$ represents the distance between the position of the $i$-th node obtained by tagged MRI and that calculated by the FEM, and $N$ is the total number of nodes.

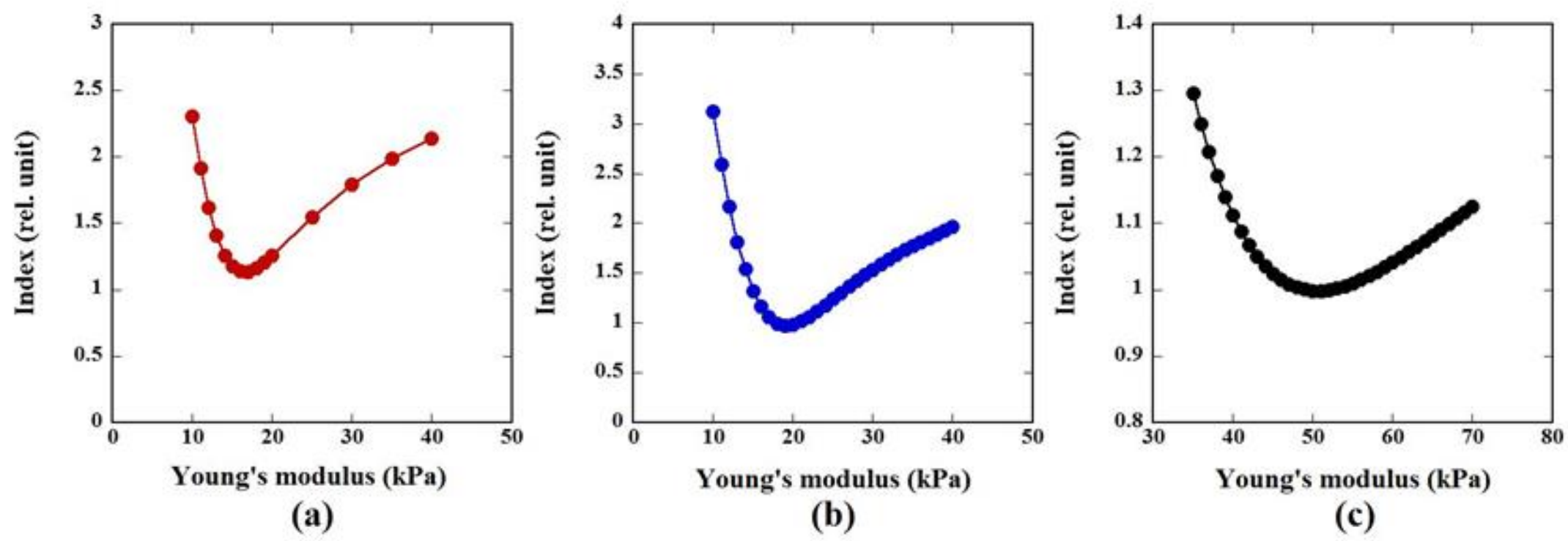

**Figure 13 Relationship between the index and Young's modulus for Phantoms A (a), B (b), and C (c). The Young's modulus was obtained from the minimum index value.**

### 3.5. Comparison of Young's moduli obtained by three methods

Table 1 summarizes the Young's moduli of Phantoms A, B, and C measured by a material testing machine, calculated by our FEM software, and obtained from tagged MRI data using the HARP method. As known from Table 1, the Young's moduli increased with increasing stiffness of phantoms in all methods. There was a tendency for the Young's moduli calculated by our FEM software and the HARP method to be slightly overestimated as compared to those measured by a material testing machine in Phantoms A and C, whereas there was a good agreement between the Young's moduli measured by a material testing machine and those calculated by the HARP method and there was a tendency for the Young's moduli calculated by our FEM software to be slightly underestimated as compared to those measured by a material testing machine in Phantom B.

**Table 1 Summary of Young’s modulus of Phantoms A, B, and C. The first, second, and third rows show the Young's moduli obtained by a material testing machine, finite element method (FEM), and harmonic phase (HARP) method, respectively. Data were represented as the mean ± SD for 5 measurements in kPa.**

| Phantom | A | B | C |
|---|---|---|---|
| Material testing | 16.6 ± 3.3 | 24.3 ± 4.0 | 36.8 ± 4.2 |

| machine | | | |
|---|---|---|---|
| FEM | 19.4 ± 4.1 | 21.6 ± 4.1 | 42.3 ± 5.3 |
| HARP | 20.4 ± 3.1 | 23.9 ± 4.0 | 41.4 ± 7.4 |

## 4. Discussion

In this study, we developed devices and software necessary for measuring the stiffness of tissues and generating its image using tagged MRI and FEM. We also performed basic studies on the feasibility and usefulness of the tissue elasticity imaging using these devices and software. It should be noted that our method is quite different from the existing method for measuring tissue stiffness using MRI, *i.e.*, MRE [6, 7].

We confirmed that the air pump developed in this study can push air out of the pump at a constant frequency without time delay and air leakage (Figs. 8 and 9). Since the intensity and frequency of the pressure can be controlled by a PC, it is possible to change their settings depending on an object. Furthermore, since the cyclic pressure device is detachable from the box shown in Fig. 3(b), it can also be used solely and applied to large organs such as the liver with the head of the cyclic pressure device being fastened with a belt.

As previously described, the strain images were generated using two different methods in this study (Fig. 1). The reason why the strain images were generated by two different methods is as follows. Although the resolution of the strain obtained by the HARP method is high, it depends on the shape of the band-pass filter used for extracting the harmonic peaks [15]. On the other hand, the strain image can also be generated from tag-cross points. Although the resolution of the strain image generated from the tag-cross points is low, the precision of the strain value obtained from the tag-cross points is high. For this reason, we used the two different strain images to generate high-resolution and high-precision strain images.

It became possible to take internal stress into consideration by combining the FEM and pressure data measured by MRI-compatible force sensors (Fig. 5). In this study, only two force sensors were used as previously described. We are now making more number of sensors, and expect that the stability and accuracy of the measurement of external force distribution will be improved by increasing the number of force sensors.

As shown in Fig. 10, the error in extracting tag-cross points was large near the boundary of the object. It may be due to the reason why our software cannot detect tags accurately in the periphery of the object in which the brightness largely changes, because our software extracts the tag-cross points by detecting the small change in the brightness. However, it was confirmed that the tag-cross points inside the object can be extracted accurately even at the last frame when the tag lines become most blurred (data not shown). Although it is necessary to confirm the extraction of tag-cross points by visual inspection and correct it manually in the present software, our software could considerably reduce the labor needed for analysis.

As previously described, calculation of Young's modulus using the FEM is based on the assumption that Young's modulus is uniform within objects. Thus, we will improve our method so that it can also be applied to the objects in which Young's modulus is not uniform. This study is currently in progress as one of the subjects for our future studies.

As shown in Table 1, the Young's moduli obtained by a material testing machine, our FEM software, and the HARP method increased with increasing stiffness of phantoms. As a whole, there were good agreements among the Young's moduli obtained by the three methods. Although the Young's moduli calculated by our FEM software were slightly overestimated in Phantoms A and C and underestimated in Phantom B as compared with those measured by a material testing machine, they were close to those obtained from tagged MRI data using the HARP method. Thus, these results suggest that our method can be applied to the measurement of Young's modulus.

## 5. Conclusion

We develop a method for tissue elasticity imaging using tagged MRI and FEM and to investigate its feasibility and usefulness using silicone gel phantoms with different degrees of stiffness. Our results suggest that our method is useful for tissue elasticity imaging and for quantifying the stiffness of tissues.

## Acknowledgements

This study was supported by a Grant-in-Aid for Scientific Research (No. 18591344) from the Japan Society for the Promotion of Science. The authors are grateful to Dr. Mitsunori Tada of National Institute of Advanced Industrial Science and Technology for the use of MRI-compatible optical force sensors.

## Appendix

The location of the harmonic peak was determined by the set values of tag space and tag angle according to Eq. (A1) [10].

$$W(t_0) = \pm \frac{2\pi}{s} \begin{bmatrix} \cos\left(\theta + \frac{\pi}{2}\right) \\ \sin\left(\theta + \frac{\pi}{2}\right) \end{bmatrix} \quad \text{(A1)}$$

where $\mathrm{W}(t_0) = [W_r(t_0) \quad W_i(t_0)]^{\mathrm{T}}$ represents the position of the harmonic peak at frame $t_0$ in a frequency domain, $s$ is the set tag space, and $\theta$ is the set tag angle. The harmonic peaks in the Fourier-transformed image include the information on the space of tag lines. When the space of tag lines becomes wider, the harmonic peaks approach a DC peak.

In this study, each harmonic peak was extracted using a band-pass filter. The shape of the band-pass filter for extracting the harmonic peak was taken as a circle with a diameter of 9 cycles/pixel, having 1 inside and 0 outside this circle. The inverse Fourier-transformed image of the band-pass-filtered harmonic peaks is a complex image. The phase image was obtained from the phase angle of the above complex image as

$$I_\varphi(t) = \arctan\left\{\frac{\mathrm{Im}[I_l(t)]}{\mathrm{Re}[I_l(t)]}\right\} \quad \text{(A2)}$$

where $I_\varphi(t)$ is a phase angle image and $I_l(t)$ is an inverse Fourier-transformed image of the band-pass-filtered harmonic peaks. Im and Re denote the imaginary and real parts of a complex number, respectively.

In general, strain ε is defined as a stretch length ($\Delta L$) divided by the original length ($L$) as

$$\varepsilon = \frac{\Delta L}{L} \quad \text{(A3)}$$

In this study, $L$ corresponds to the set tag space. The strain at frame $t$ [$\varepsilon(t)$] is given by

$$\varepsilon(\mathrm{t}) = \frac{s(t) - s(t_0)}{s(t_0)} \quad \text{(A4)}$$

where $s(t_0)$ and $s(t)$ represent the tag space at frame $t_0$ and $t$, respectively.

The derivative of the phase angle at frame $t$ given by Eq. (A2) with respect to the position $y$ can be given by

$$\frac{\partial I_{\varphi}(t)}{\partial y} = \frac{2\pi}{s(t)} \tag{A5}$$

Substituting Eq. (A5) into Eq. (A4) yields

$$\varepsilon(t) = \frac{\frac{\partial I_{\varphi}(t_0)}{\partial y}}{\frac{\partial I_{\varphi}(t)}{\partial y}} - 1 \tag{A6}$$

It should be noted that since $I_{\varphi}(t)$ ranges from $-\pi$ to $\pi$ because of the inverse tangent operator given by Eq. (A2), the phase angle exceeding $\pi$ suddenly moves to $-\pi$. Such a sudden transition in phase angle from $\pi$ to $-\pi$ is called "phase wrap". Thus, it is necessary to correct for such a sudden transition in phase angle, *i.e.*, "phase wrap correction".

Figure 14 illustrates the procedure for obtaining the derivative of phase angle [$\partial I_{\varphi}(t)/\partial y$ given by Eq. (A5)] and phase wrap correction. First, $\partial I_{\varphi}(t)/\partial y$ was calculated from $I_{\varphi}(t)$ using a forward difference method. Since $\partial I_{\varphi}(t)/\partial y$ always has a positive value, the points at which $\partial I_{\varphi}(t)/\partial y$ had a negative value were linearly interpolated using two points of the neighbor as shown in the fifth row of Fig. 14. We call this method "the linear-interpolation method" in this study.

As shown in Fig. 10(b), the harmonic peak is symmetrical about the origin. In this study, we extracted two symmetrical harmonic peaks, corrected for phase wrap, and calculated the gradients of phase angles separately. After these procedures, we averaged these two wrap-corrected gradients of phase angles. Note that since the inverse Fourier-transformed image of each symmetrical peak has the reverse sign of imaginary term, we calculated the average of wrap-corrected gradients of phase angles after the sign of one of them was reversed.

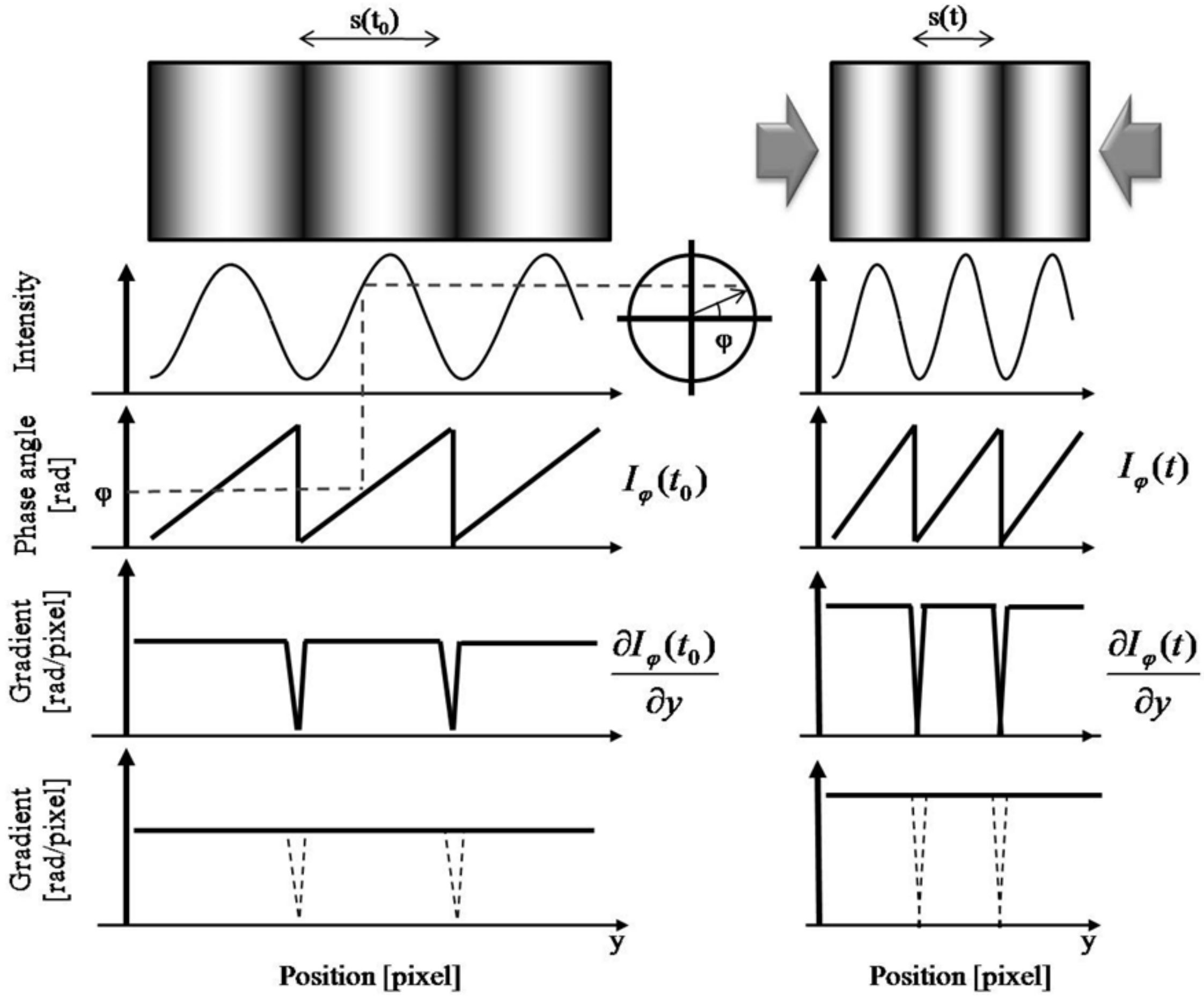

**Figure 14 Illustration of the procedure for obtaining $\partial I_{\varphi}(t)/\partial y$ and phase wrap correction. The procedure at the first frame, in which no pressure is applied, is shown on the left panels, while that at frame $t$, in which pressure is applied, is shown on the right panels. The first, second, third, fourth, and fifth rows show simple models of tagged MRI, signal intensities of tagged MRI, phase angles, gradients of phase angles, and wrap-corrected gradients of phase angles as a function of position, respectively.**